\documentclass[prl,aps,twocolumn,superscriptaddress,showpacs]{revtex4}

\usepackage{epsfig}
\usepackage{amsmath}
\usepackage[mathscr]{eucal}

\newcommand{\be}{\begin{equation}}
\newcommand{\ee}{\end{equation}}
\newcommand{\bea}{\begin{eqnarray}}
\newcommand{\eea}{\end{eqnarray}}

\newcommand{\p}{\partial}

\newcommand{\e}{\ensuremath{\mathrm{e}}}
\newcommand{\HH}{\ensuremath{\mathscr H}}

\newcommand{\ket}[1]{\ensuremath{\vert{#1}\rangle}}

\newcommand{\im}{\ensuremath{\mathrm{i}}}
\newcommand{\tm}{\ensuremath{t_{\mathrm{m}}}}
\newcommand{\VSD}{\ensuremath{V_{\mathrm{SD}}}}

\usepackage{grffile}

\usepackage{ulem}
\usepackage{color}
\definecolor{darkblue}{rgb}{0.0,0,0.6}
\definecolor{darkred}{rgb}{0.6,0,0.0}

\begin{document}

\title{Transport through nanostructures: Finite time vs.\ finite size}

\author{Peter Schmitteckert}
\affiliation{DFG Center for Functional Nanostructures, Karlsruhe Institute of Technology, 76128 Karlsruhe, Germany}
\affiliation{Institute of Nanotechnology, Karlsruhe Institute of Technology, 76344 Eggenstein-Leopoldshafen, Germany}

\author{Sam T. Carr}
\affiliation{School of Physical Sciences, University of Kent, Canterbury CT2 7NH,
UK}
\affiliation{DFG Center for Functional Nanostructures, Karlsruhe Institute of Technology, 76128 Karlsruhe, Germany}

\author{Hubert Saleur}
\affiliation{Institut de Physique Th\'eorique, CEA, IPhT and CNRS, URA2306, Gif Sur Yvette, F-91191}
\affiliation{Department of Physics, University of Southern California, Los Angeles, CA 90089-0484}

\date{\today}

\pacs{73.63.-b, 72.70.+m, 05.40.Ca, 05.60.Gg} 

\begin{abstract}
Numerical simulations and experiments on nanostructures out of equilibrium  usually exhibit strong finite size and finite measuring time $\tm$ effects. We discuss how these affect the determination of the 
full counting statistics  for a general quantum impurity problem. We find that, while there are many methods available to improve upon finite-size effects, any real-time simulation or experiment will still be subject to finite time effects: in short size matters, but time is limiting.
We show that  the leading correction to the cumulant generating function (CGF) at zero temperature for single-channel quantum impurity problems goes as $\ln \tm$ and is universally related to the steady state CGF itself for non-interacting systems. We then give detailed numerical evidence
for the case of  the self-dual interacting resonant level model that this relation survives the addition of interactions.  This allows the extrapolation of finite measuring time in our numerics to the long-time limit, to excellent agreement with Bethe-ansatz results.
\end{abstract}

\maketitle

Finite (measuring) time effects play a crucial role in the analysis of nanostructures, in particular of their out of equilibrium properties \cite{Schoenhammer-2007}. 
This is clear for numerical simulations,  where, rather than solve an equilibrium eigenvalue problem, one must now time evolve from the non-equilibrium initial condition:
\begin{equation}
    \ket{\Psi(t)} = \e^{-\im\hbar \HH(t-t_0)} \ket{\Psi(t_0)} . \label{eq:TE}
\end{equation}
Time evolving a many-body state is a computationally expensive procedure, which results in a limit on the time-scale  computationally accessible.
The same may be true of real experiments, where similarly one can't measure the system for eternity.  
Indeed in recent experiments concentrating on the Full Counting Statistics (FCS) \cite{Gustavsson-2006,Gustavsson-2007,Choi-2012}, the bias voltage is so low that the relevant parameter, the dimensionless measuring time $\VSD\tm$, is actually rather small.  

Finite time effects are often combined with finite size effects, which are ever present  for systems on nanoscale structures. In  equilibrium, finite size scaling is well
under control, and, interestingly,  often turns out to encode fundamental properties of the bulk system. For example the relation between the $1/L$ ($L$ being the system size) corrections to the ground state energy density and the central charge of the system \cite{Bloete-1986,Affleck-1986}. 

By analogy with the situation of the finite size, one may therefore ask the questions: Can one extrapolate short finite measuring time $\tm$ results to the long time limit, is there any univesality in these finite $\tm$ corrections, and if so, can these corections give us new information about the system?
In this Letter, we will show that in certain situations the answer to all three of these questions in the case of the FCS is yes.  This will require us to  also  disentangle the  contributions  that come from \textit{finite size} and \textit{finite measuring time}.  

The systems we  consider  are quantum impurities coupled to two non-interacting leads which are (initially) held at different chemical potentials.  The finite size in question, $L$, is then the size of the lead, the quantum impurity being naturally a small size.  The plan for the rest of the paper is  to introduce the FCS, look at the intrinsically finite size corrections, then turn to the major issue of the work which is how to understand the intrinsically finite time effects.

The  transport properties  of  a nanostructure are of course not entirely encoded in the average current $\bar{I}$ flowing for a given bias voltage $\VSD$ --  fluctuations are of crucial importance, for example in the determination of the charge of the carriers.  The corresponding  information is conveniently assimilated within the framework of FCS.   In the traditional two-terminal setup, one studies the probability distribution $P_{\tm}(n)$ that a charge $Q=ne$ has been transferred from the left to the right lead in the measuring time $\tm$ ($e$ being the fundamental charge on the electron) \cite{LL,Levitov-Lee-Lesovik-1996,Nazarov-Blanter-book,Belzig-lecture,Bagrets-Nazarov-2003,Klich-2003,Komnik-2007}. 
Rather than working directly with the distribution, it is usually more informative to study the cumulant generating function (CGF), defined as \cite{LL,Levitov-Lee-Lesovik-1996}
\begin{equation}
F_{\tm}(\chi) = -\ln \left[\sum_n e^{\im n\chi} P_{\tm}(n)\right]. \label{eq:CGF}
\end{equation}
The irreducible cumulants of charge transfer are then simply obtained via $C_n = \left. - \left( \frac{\p}{\im\p \chi}\right)^n F(\chi) \right|_{\chi=0}$, while the periodicity of the CGF yields information regarding the charge of the quasi-particles involved in transport \cite{Belzig-lecture,Saleur-Weiss-2001,Levitov-Resnikov-2004} which in a strongly correlated system may not be simple electrons, and may even undergo a change as a function of bias voltage \cite{CBS-2011,Ivanov-Abanov-2010}.  
In the long time limit, each of the cumulants (and by inference, the CGF) is proportional to the measuring time $\tm$, for example the first cumulant gives the current $C_1\sim \bar{I} \tm$, while the second $C_2\sim \bar{S}\tm$, where $\bar{S}$ is the zero-frequency shot-noise. 

 In the present work, we will be particularly interested in the leading corrections to these expressions for finite measurement times.  However, we first discuss the corrections due to the finite size of the leads.  To make this discussion concrete, we focus on the interacting resonant level model (RLM), \cite{CBS-2011,Boulat-Saleur-2008, Boulat-Saleur-Schmitteckert-2008, Branschaedel_Boulat_Saleur_Schmitteckert:PRL2010,Branschaedel_Boulat_Saleur_Schmitteckert:PRB2010,Andergassen-2011,Kennes-Meden-2013,Doyon-2007,Bernard-Doyon-2012,Genway-2012} described by the Hamiltonian
 \begin{multline}
 {\mathscr H} = \sum_{n=L,R} \left\{ -\sum_{i=0}^{M_n} (c_{n,i}^\dagger c_{n,i} + H.c.) 
 + J' c_{n,0}^\dagger d + H.c. \right.
   \\ \left. + U(d^\dagger d - 1/2)(c_{n,0}^\dagger c_{n,0} -1/2) \right\}. \label{eq:IRLM}
 \end{multline}
In this expression, $d^\dagger$ creates a fermion on the resonant level, while $c_{n,i}^\dagger$ creates a fermion on the right or left lead at site $i$, the total size of each lead being $M_{L,R}$.  The hybridization between the leads and the resonant level is $J'$ (the hopping parameter on the leads which sets the overall energy scale of the problem has been set to $1$), and $U$ gives an interaction between the resonant level and the leads.

There are two values of $U$ where the model has been solved out of equilibrium and transport properties are known: $U=0$ is the non-interacting case \cite{Branschaedel_Boulat_Saleur_Schmitteckert:PRB2010,Bernard-Doyon-2012,Genway-2012}, and $U=2$ where the model shows a certain duality \cite{CBS-2011,Boulat-Saleur-2008, Boulat-Saleur-Schmitteckert-2008, Branschaedel_Boulat_Saleur_Schmitteckert:PRL2010} although many works exist on the model at more general values of interaction \cite{Andergassen-2011,Kennes-Meden-2013,Doyon-2007}.  
We imagine a situation where the two leads are initially decoupled from the resonant level, and exhibit a charge imbalance characterized by a difference in potential $\VSD$, modelling the source-drain potential in an experimental setup.  
At time $t=0$, the coupling between the leads (through the resonant level) is quenched on, and a current begins to flow \cite{PS:PRB04,PS:Ann2010}; an extension of this method to calculate the CGF of FCS was recently expounded in \cite{CBS-2011}.  
We refer to the literature for a physical discussion of the transport properties of the RLM: here we use it as a basis for discussions of the effects on transport of the finite size $M_{L,R}$ of the leads, and the finite measuring time $\tm$ on the FCS.  We expect our results to be applicable to more generic quantum impurity models.

One finds that there are three important consequences to having finite size leads.  The first concerns the discrete nature of the energy levels of the leads, leading to a finite-size energy gap, $\Delta$.  This means that all important physical processes must happen at energy scales larger than this gap.
This is a phenomenon inherited from equilibrium problems, which remains relevant to the present non-equilibrium case.  The second consequence also relates to the finite-size energy gap but now is intrinsic to the transport -- a coherent system with a gap exhibits oscillations in the DC transport. These were first observed in the current \cite{Boulat-Saleur-Schmitteckert-2008} but are also seen in all higher cumulants (or alternatively the CGF) \cite{CBS-2011}.  Even in systems when the finite-size gap is unimportant for equilibrium properties, the oscillations $\sim \Delta \cos\left( \VSD \tm + \eta \right)$ may be clearly visible in non-equilibrium DC transport. 
These oscillations do not decay in time, and have the same physical origin as the Josephson effect, with the frequency  given by the source drain voltage $\VSD$ and the amplitude proportional to the gap \cite{Boulat-Saleur-Schmitteckert-2008,PS:Ann2010}, i.e.\ proportional to $1/L$ if the leads are discretized uniformly in energy space. 
Furthermore, so long as one evolves in time for sufficiently long to see a few oscillations, results may be fitted using the above expression; this procedure has been remarkably successful provided the bias $\VSD$ is not too small.
The third consequence is that after a transit time $t_{\mathrm T}= v_c L$ (where $v_c$ is the excitation velocity in the leads) the excitations leaving one lead bounce off the edge of the other lead, which eventually causes the current to flow the other way. For details we refer to \cite{PS:Ann2010}.  

\graphicspath{{./Figures/}}
\begin{figure}
\begin{center}
\includegraphics[width=\columnwidth,clip=true]{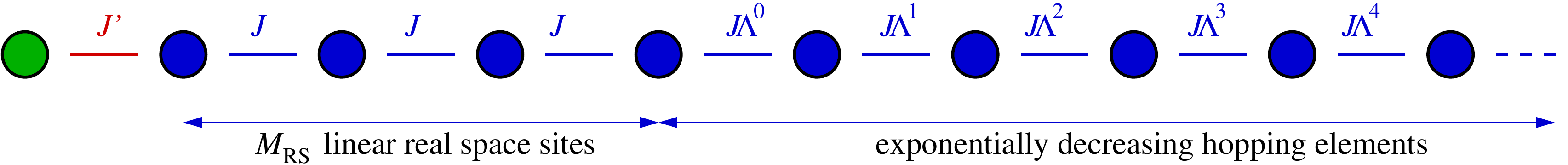}\vspace{-0.5cm}
\end{center}
\caption{[Color online] Setup of damped boundary conditions for a resonant level coupled to a single lead via a hybridization of J'.
         Each lead is first described by a homogeneous tight binding chain, which is then followed by  exponentially decreasing 
          hopping elements in an NRG fashion 
         in order to increase the density of states at the Fermi surface.
        }\label{fig:DBC}
\end{figure}

In certain physical systems, including the (interacting) RLM, the finite-size gap is relatively harmless.  However, in certain strongly correlated systems, epitomized by the Kondo effect, the interesting emergent phenomena occur at low energy scales.  In these cases 
$\Delta$
must be smaller than any physical
scale in which one is interested.
For impurity problems in equilibrium this problem may be solved via numerical renormalization \cite{Wilson:NRG}
 which has turned
out to be one of the most important tools for equilibrium strongly correlated systems \cite{Pruschke_Costi:NRG}.
One first introduces a logarithmic discretization of the leads  in energy space, which is then
transformed into a nearest-neighbour tight binding chain with exponentially decreasing hopping elements, leading
to an exponentially enhanced density of states close to the Fermi surface.  This approach has also been
 extended to non-equilibrium systems \cite{Anders_Schiller:PRL2005} within
a time dependent NRG (td-NRG) method. In this case, one first solves the non-interacting scattering problem,
one then discretizes the resulting scattering states in analogy to equilibrium NRG, and finally one switches on interactions
perturbatively.
The concept of increasing energy resolution by changing the bond terms was extended to smooth boundary conditions \cite{Vekic_White:PRL1993} 
in the context of bulk systems and to damped boundary conditions (DBC) in \cite{Bohr_PS:PRB2007,PS:JPCS2010}, where a homogeneous tight binding chain is
inserted between the impurity and the exponentially damped region, see Fig.~\ref{fig:DBC}.

While this setup proved to be successful for  the linear conductance of the IRLM \cite{Bohr_PS:PRB2007}
it turned out to be problematic for non-equilibrium properties.
In \cite{PS:JPCS2010} it was shown that in time dependent simulations the exponentially decreased hopping elements lead
to an exponentially decreased excitation velocity in the damped region, resulting into an NRG tsunami:  the leads
lose the property of a nicely behaved bath (see also \cite{Rosch:2012}). In addition, each link with changed hopping elements acts as an
additional scatterer leading to an increased backscattering.

\begin{figure}
\graphicspath{{./Figures/}}
\begin{center}
\includegraphics[width=\columnwidth,clip=true]{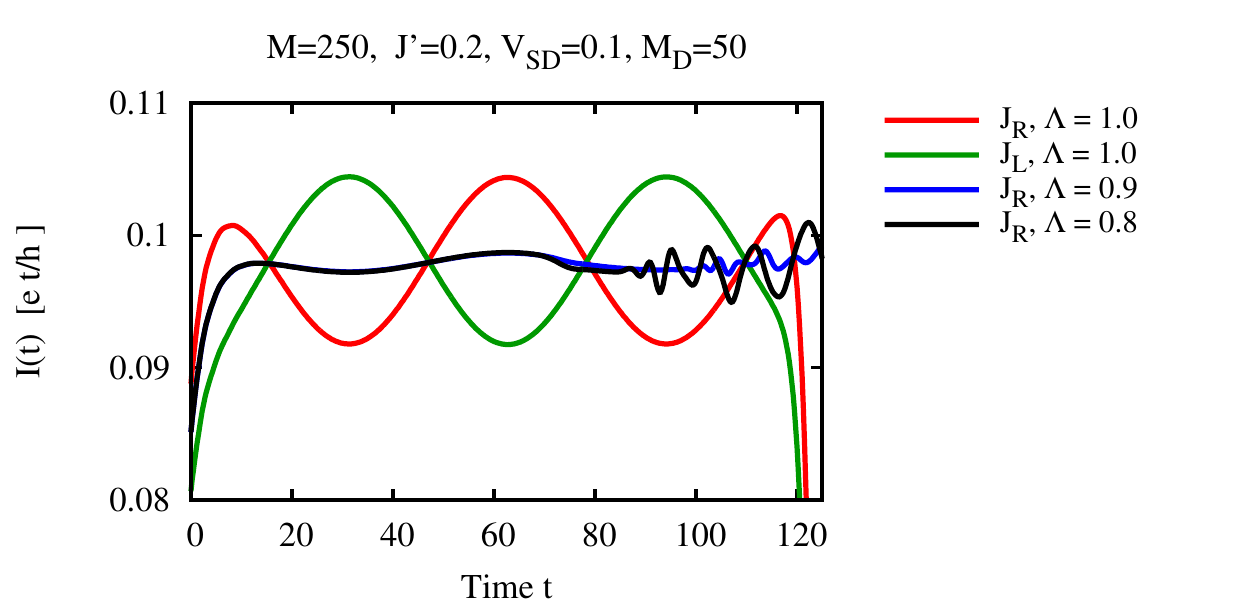}\vspace{-0.5cm}
\end{center}
\caption{[Color online] Time evolution of the current following a charge imbalance quench 
      measured to the left (L) and right (R) of the resonant level in model \eqref{eq:IRLM} with $U=0$ and $J'=0.2$.  The system size here is $M=250$ sites. The red and green lines
      displaying the large oscillations correspond to homogeneous leads.
      The blue and black line corresponds to damped leads with $\Lambda=0.9,0.8$ in the setup shown in Fig.~\ref{fig:DBC}}\label{fig:JO}
\end{figure}

In Fig.~\ref{fig:JO} we show the current as a function of time following the charge imbalance quench (using the general protocol as described above) for the non-interacting RLM with $J'=0.2$ in a system with DBC, where the hopping elements on the last 50 sites
of the left and right leads are decreasing with a factor of $\Lambda=1.0$ (homogenous leads), $0.9$, and $0.8$. The $\Lambda=1$
displays the large Josephson oscillations (JO) as discussed above until the transit time $t_{\mathrm T}$ when we
see the back reflection of the hard wall boundaries of the leads.  It is worth noting that for this particular example though, there is a phase
shift of $\pi$ between the response left and right of the impurity; averaging over these dramatically decreases the size of the oscillations.

As one may expect, the DBC lead to a decreased
height of the JO. However, the gain is not exponentially large, as we are now at finite voltage $\VSD=0.1J$,
while the DBC lead to exponentially enhanced DOS at the Fermi surface only.  One could improve by using
a different discretization scheme where the enhancement of the DOS is shifted to energies $\pm \VSD$ \cite{PS:JPCS2010}.
However one still faces the problem that each modified bond will still lead to a back reflection.
The transit time for the DBC systems is decreased compared to the homogeneous case by $2 M_{\mathrm D}/ v_c$,
where $M_{\mathrm D}$ is the number of modified bods.  Accordingly, we have  wiggles appearing shortly 
after $t_{\mathrm T}= (M-2M_{\mathrm D})/v_c$. In principle, by using reflectionless DBC \cite{Moliner_PS:EPL2011}
one could avoid these wiggles, however a reduced transit time remains.

We now say a few words about this issue of transit time, which places a hard limit on the length of time one may evolve the system before finite-size effects interfere with the time evolution.  Here also, one can minimize this disruption (but only in a homogenous lead) by using a \textit{conformal time}.  In equilibrium,  it is well known how conformal invariance allows one, via a  finite size/temperature transformation, to control the effects of finite imaginary time \cite{CFT-book}. The generalization to out of  equilibrium situations and real time amounts to replacing 
\begin{equation}
 \tm\rightarrow   d(\tm) = \left( \sin\frac{ \pi \tm }{M/v_c} - \sin\frac{ \pi t_0 }{M/v_c} \right) \frac{ M \pi}{v_c} \,, \label{eq:ConformalTime}
\end{equation}
where $t_0$ is the initial time after quenching the system at $t=0$ until the counting field is switched on at $t=t_0$. 
While for times much less than the transit time $d(t) \approx t$, as one approaches the transit time, 
the above formula captures the leading effects of the back-reflection from the leads remarkably well (despite  not being entirely justified theoretically \cite{Cardy-Calabrese-2006}) -- see supplementary material for examples.

While such consideration shows that there are intimate connections between finite size and finite time effects, computer constraints means that one may not always be able to time evolve the system as long as the transit time \cite{DMRG-runaway}.  Furthermore, even if one finds an ideal method that works in non-equilibrium situations to eliminate the finite size effects, any real-time numerical simulation will nevertheless have to be cut off after some finite running time.  We therefore now turn to effects intrinsic to the finite measuring time of the system, and show that  these may be much larger corrections than any of those directly due directly to the finite size. 
To demonstrate this, we look at the CGF of the FCS as a function of  $\tm$.
 
As mentioned previously, one expects the cumulants -- and by extension the CGF \eqref{eq:CGF} -- to grow linearly in measuring time.  As zero temperature,
the subleading corrections are of logarithmic nature \cite{Muzykantskii-2003,Hassler-2008}
\begin{align}
		 F(\chi, \tm ) &=\; \tilde{F}_0\tm \,+\, \tilde{F}_1 \ln\left(\VSD\tm\right) \,+\, \cdots \nonumber \\
	\implies \,	 \dot F(\chi, \tm ) &=\; \tilde{F}_0 \,+\, \tilde{F}_1/\tm \,+\, \cdots \label{eq:series}
\end{align}
Formally, this is an expansion of the CGF in the small parameter $(\VSD\tm)^{-1}$.  The long measuring time limit, $\tilde{F}_0$, is what is commonly quoted and analyzed as the FCS, and is given for non-interacting particles (at zero temperature) by the Levitov-Lesovik formula \cite{LL,Levitov-Lee-Lesovik-1996}.

Here, we \textit{conjecture} that the leading correction to this, $\tilde{F}_1$, is independent of the quench protocol (i.e. is a true steady state property), and given in the zero temperature limit by
\begin{equation}
\tilde{F}_1 = \frac{1}{\pi} \left( \frac{d \tilde{F}_0}{d\VSD} \right)^2. \label{eq:F1}
\end{equation}

Eq.~\eqref{eq:F1} is valid for single-channel quantum impurity problems, for systems symmetric with respect to the sign of the applied voltage.
While it may only be formally derived for non-interacting fermions, we will present arguments and numerical evidence that suggest it survives the addition of interactions.  However we stress that a proof of this equation, or alternatively an understanding of the limits of its applicability, is an open question.

Eq.~\eqref{eq:F1} agrees with previously derived results for non-interacting fermions \cite{Muzykantskii-2003,Hassler-2008,Schoenhammer-2007}, where the only essential feature that goes into deriving this term is the Fermi-edge singularities [see supplementary material].  Thus the result is limited to zero temperature, however in this limit one expects it to be universal as it does not involve details of the quench.  For small but non-zero temperatures $T$, we would expect the result still to hold on time scales $\tm<1/T$ \cite{Braunecker-2006}, at later times corrections to the long-time limit no longer being universal.  However this is beyond the scope of the present work.

The fact that the physics of $\tilde{F}_1$ is limited to the Fermi-edge, which is explicitly captured by the derivative representation Eq.~\ref{eq:F1},  gives hope that this formula may also be valid in the interacting case.  In a nearly-free electron picture where interactions may be treated as perturbatively dressing free-electron results, one would certainly imagine that the relationship remains unchanged between $\tilde{F}_0$ which involves all states within an energy window of width $\VSD$, and $\tilde{F}_1$ which involves only the states at the Fermi edges.  This can be made more formal by looking at the second cumulant (shot noise), where the correction according to Eq.~\eqref{eq:F1} is $\propto G^2$, $G$ being the differential conductance.  The finite-time correction to the second cumulant of FCS is directly related to the finite-frequency correction to shot noise, something that has also been investigated in detail for the third cumulant \cite{Salo-2006}.  
This relation can therefore be compared to an earlier conjecture that the frequency dependent noise $S(\omega)-S(0) \propto G^2 |\omega|$.   
In \cite{Chamon-Freed-1999}, this was shown to be true to all orders in perturbation theory, while other work \cite{Lesage-Saleur-1997} suggested that this may break down in a non-perturbative regime.  
This question has been revisited recently numerically \cite{Branschaedel_Boulat_Saleur_Schmitteckert:PRL2010,Branschaedel_Boulat_Saleur_Schmitteckert:PRB2010} which supported the idea that this simple relation holds even non-perturbatively.  

As one final comment on the conjecture \eqref{eq:F1}, we note that it tells us that universal finite-measuring time corrections are absent for the first cumulant (current).  This must hold if the notion of a steady state has a meaning.

While at present we are not able to give a more substantive analytic derivation of the conjecture \eqref{eq:F1}, we now back it up with numerical evidence.
In the supplementary material, we demonstrate the numerical procedure with the non-interacting RLM.  Here we apply a more stringent test, using
 the self-dual interacting RLM.  This is chosen as it exhibits non-trivial correlations, and is one of few such models where exact results for the FCS (in the long-time limit) are known analytically \cite{CBS-2011}.  For convenience, the analytic results are given in the supplementary online material.
In Fig.~\ref{fig:IRLM} we compare the real part of  ${\dot F}$ obtained numerically with the analytic result including the $1/\tm$ correction, assuming
that Eq.~\eqref{eq:F1} holds.  As can be seen there is a nice agreement over four orders of magnitude.
Although there appears a shift in each curve we would like to stress that simulations are done on a lattice,
while the analytic results are taken from a continuum  theory and the only scale parameter linking the two is taken from
previous work \cite{Boulat-Saleur-Schmitteckert-2008}.  Similar agreement can be seen at other values of $\VSD$ or $J'$.

\begin{figure}
\begin{center}
\includegraphics[width=\columnwidth,clip=true]{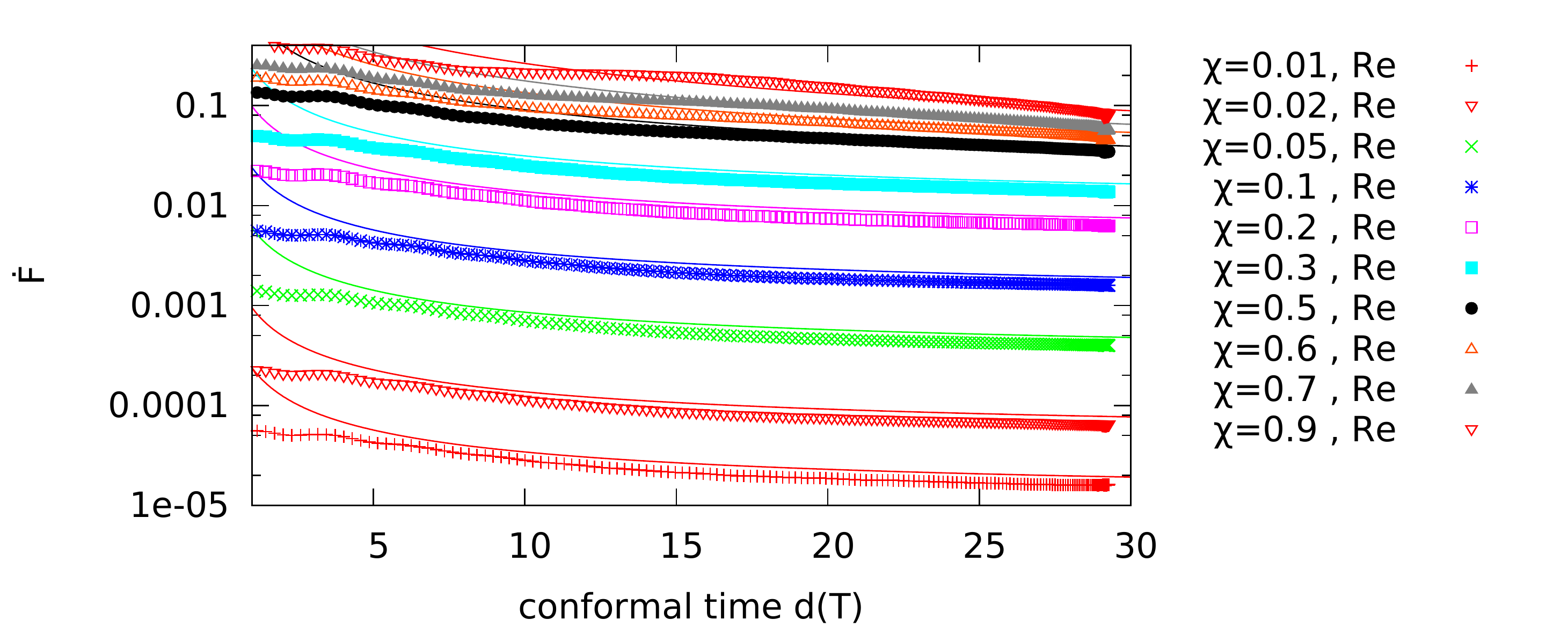}\vspace{-0.5cm}
\end{center}
\caption{[Color online] Real part of  $\dot{F} $ for the SD IRLM at $\VSD=0.3$, $J'=0.2$.
	Symbols correspond to the numerical result, the lines are the  analytical results given in the supplementary material.
    The system size is $M=240$ sites. } \label{fig:IRLM}
\end{figure}

\begin{figure}
\begin{center}
\includegraphics[width=\columnwidth,clip=true]{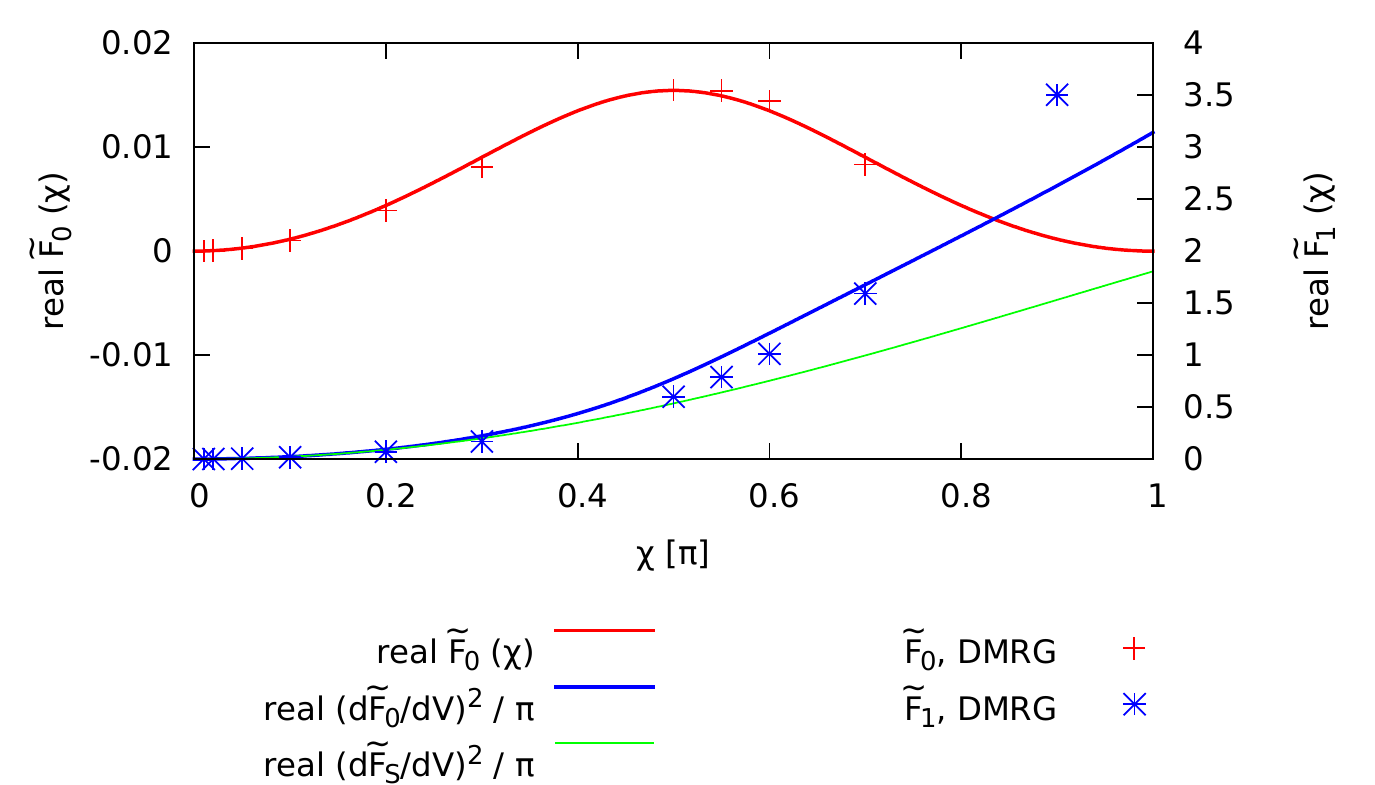}\vspace{-0.5cm}
\end{center}
\caption{ [Color online] Comparison of the analytical and numerical results of the leading $\tilde{F}_0$ and subleading $\tilde{F}_1$ contributions to the CGF of the self-dual interacting RLM.  The numerical results are obtained from fitting the real-time data in Fig.~\ref{fig:IRLM}, the analytic expressions are given in the supplementary material.  
} \label{fig:IRLMcomp}
\end{figure}

Using these results we then perform a fit as a function of $\tm$ to the series \eqref{eq:series}.  To avoid influence of the transients, 
we limit the fit to $d(\tm)>13$.
These are compared to the analytic results in Fig.~\ref{fig:IRLMcomp}.  One sees in particular a very good agreement for $\tilde{F}_1$ until $\chi$ becomes too large, where the numerical data is very messy for reasons not yet fully understood.  This gives excellent evidence in support of the conjecture \eqref{eq:F1}.  We also plot for comparison the quadratic in $\chi$ approximation to $\tilde{F}_1$ -- which is the term coming from the previously discussed correction to the shot noise.  It is clear from the plot that both the true $\tilde{F}_1$ from \eqref{eq:F1} and the numerical data deviate significantly from the quadratic approximation -- in other words, we are seeing beyond the lowest cumulants.

It is also worth noting the difference in scale for the $\tilde{F}_1$ and $\tilde{F}_0$ plots -- the finite $\tm$ correction is much larger than the long-time limit until $\tm\sim 100$.  Nevertheless, a correct fitting procedure as a function of measurement time allows one to extrapolate over several orders of magnitude and see (to good agreement with the analytic result) the bump in the long-time CGF $\tilde{F}_0$; a feature that was entirely absent in the previous analysis \cite{CBS-2011}.

In summary, we have discussed how  dc transport calculations are subject to finite time and finite size effects
and both are of different nature. While finite size effects can be controlled by a suitable choice
of boundary conditions, the situation for finite time effects is more difficult. 
By looking at the cumulant generating function one can perform a systematic extrapolation towards the long time limit.
In addition we showed that in the examples given the leading finite time corrections 
are intricately related to the long-time CGF itself,  see Eq.~\eqref{eq:F1}.  If such a relation holds more generally, it provides
a fantastic possibility for self-consistency checks within time-dependent simulations.  We therefore hope that this Letter stimulates further work supporting or disproving a more general validity of \eqref{eq:F1}.

We would like to thank Andreas Komnik, Dmitry Bagrets, and Dmitry Gutman for insightful discussions.




\begin{widetext}

\section{Online Supplementary Material}

\subsection{FCS in non-interacting systems}

The leading term in the long-time limit of the CGF for non-interacting Fermions is given by the well known Levitov-Lesovik formula \cite{sm:Levitov-Lee-Lesovik-1996}, which at zero temperature $T=0$ reads
\begin{equation}
\tilde{F}_0(\chi) = \int_{-\VSD/2}^{\VSD/2} d\epsilon \ln \left(1+T(\epsilon)(\e^{\im\chi} - 1 ) \right). \label{eq:LL}
\end{equation}
Here $T(\epsilon)$ is the transmission through the quantum impurity at energy $\epsilon$.

The sub-leading correction in measuring time is then given by \cite{sm:Hassler-2008}
\begin{align}
\tilde{F}_1(\chi) = \frac{1}{2\pi} \sum_{\epsilon=\pm \VSD/2} \ln^2 \left(1+T(\epsilon)(\e^{\im\chi} - 1 ) \right), \label{eq:NI-F1}
\end{align}
which is seen to agree with the more general conjecture \eqref{eq:F1}, so long as $T(\epsilon)=T(-\epsilon)$.  If the transmission is not symmetric around the chemical potential, one can easily generalize \eqref{eq:F1} -- the result would now depend on the derivatives with respect to applied biases on each lead.

This result for non-interacting Fermions was first derived in 
 \cite{sm:Hassler-2008} using a wave-packet approach; it is also is in accordance with alternative studies that were limited to energy-independent transmissions \cite{sm:Muzykantskii-2003,sm:Schoenhammer-2007}.
It is instructive however to understand fundamentally where the term $\tilde{F}_1$ logarithmic in measuring time originates from.  Using the Klich representation, \cite{sm:Klich}, one sees that the CGF may be written as a determinant which is of Toeplitz form.  The leading contribution to the CGF is then given by Szego's theorem \cite{sm:Szego}, which is the Levitov-Lesovik result \cite{sm:Hassler-2008}.  However, one knows from the Fisher-Hartwig conjecture \cite{sm:FH} that if the matrix elements contain jumps, which in this case comes from the Fermi-edges at zero temperature, then the Toeplitz determinant acquires additional logarithmic corrections, which gives the result \eqref{eq:NI-F1}.  Similar ideas have also recently been applied to Luttinger liquids \cite{sm:Dima1,sm:Dima2,sm:Dima3}.


\subsection{$\tilde{F}_0$ and $\tilde{F}_1$ for the non-interacting resonant level model}

The transmission through the non-interacting RLM \eqref{eq:IRLM} with $U=0$ is easy to calculate (see e.g.~\cite{sm:Branschaedel_Boulat_Saleur_Schmitteckert:PRB2010}) and yields
\be
T(\epsilon) = \frac{1-a^2\epsilon^2}{1+b^2\epsilon^2} \label{eq:T-RLM}
\ee
where $a^2 = 1/4$ and $b^2=\frac{1-2J'^2}{4J'^4}$.  Notice that this is relevant for the cosine-band of the lattice model \eqref{eq:IRLM}; in our units the more usual wide-band limit corresponds to $\epsilon,J' \ll 1$ and yields the usual Lorentzian form $T(\epsilon)=1/(1+\epsilon^2/\Gamma^2)$ where $\Gamma=2J'^2$.

Executing the integral in Eq.~\eqref{eq:LL} gives us
\be
\tilde{F}_0(\chi) = \VSD\ln \left[ 1+ \frac{(e^{\im\chi}-1)(1-a^2\VSD^2/4)}{1+b^2\VSD^2/4}\right] + \frac{4e^{\im\chi/2}\tan^{-1}\left[\frac{\VSD}{2}e^{-\im\chi/2}\sqrt{b^2-a^2(e^{\im\chi}-1)}\right]}{\sqrt{b^2-a^2(e^{\im\chi}-1)}}- \frac{4\tan^{-1}(b\VSD/2)}{b}. \label{eq:F0-RLM}
\ee
Simple substitution of \eqref{eq:T-RLM} into \eqref{eq:NI-F1} gives 
\be
\tilde{F}_1(\chi)=\frac{1}{\pi}  \ln^2 \left(1+ \frac{1-\VSD^2/16}{1+\frac{1-2J'^2}{16J'^4}\VSD^2}  (\e^{\im\chi} - 1 ) \right),\label{eq:F1-RLM}
\ee

\subsection{Numerical results for non-interacting resonant level model}

As previous numerical studies of the evolution of the CGF as a function of measuring time have concentrated only on the case of transmission independent of energy \cite{sm:Schoenhammer-2007}, we include here results for the non-interacting resonant level model.  The expected analytic expressions for $\tilde{F}_0$ and $\tilde{F}_1$ are obtained above in Eqs.~\eqref{eq:F0-RLM} and \eqref{eq:F1-RLM}.

\begin{figure*}
\begin{center}
\includegraphics[width=0.49\columnwidth,clip=true]{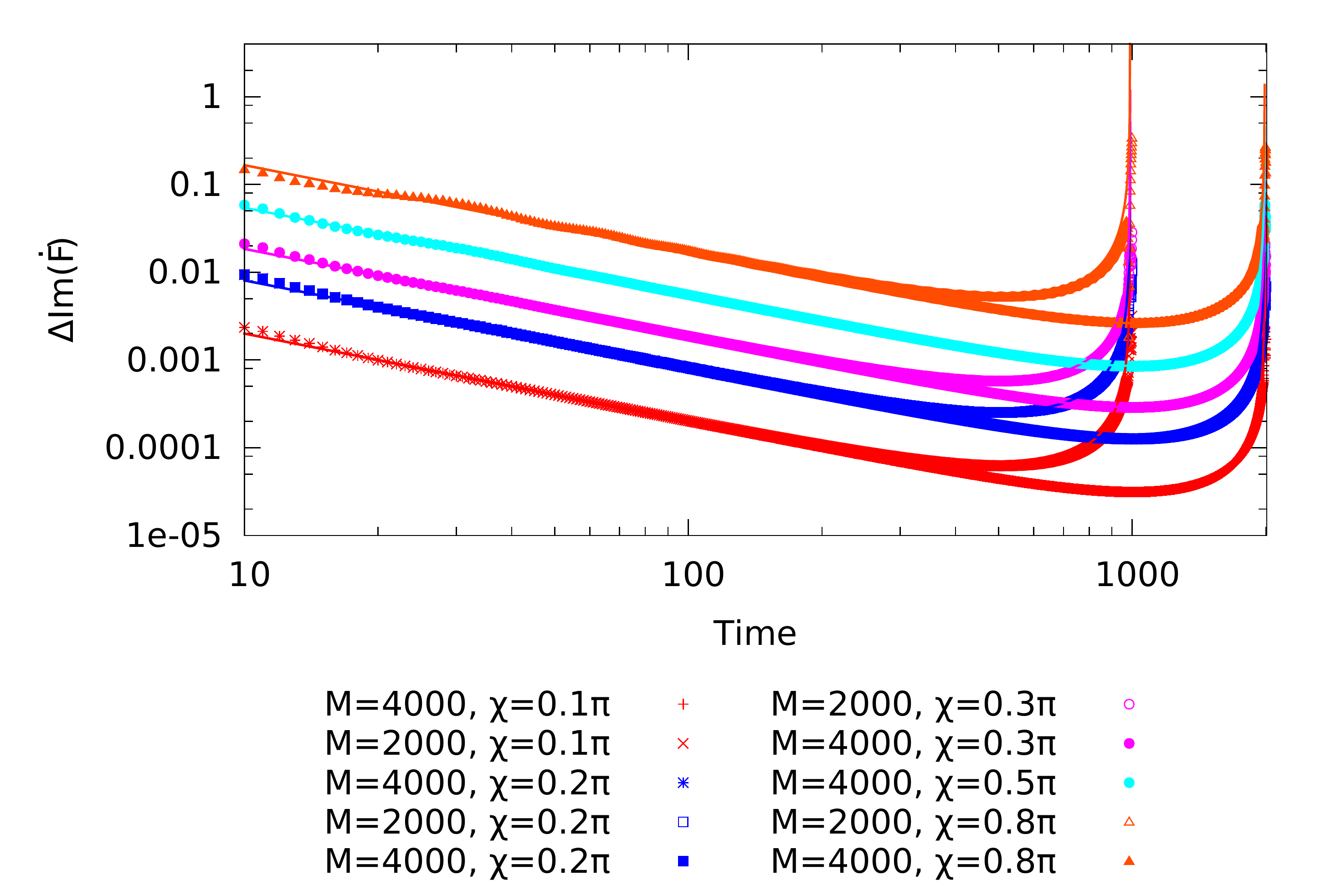}
\includegraphics[width=0.49\columnwidth,clip=true]{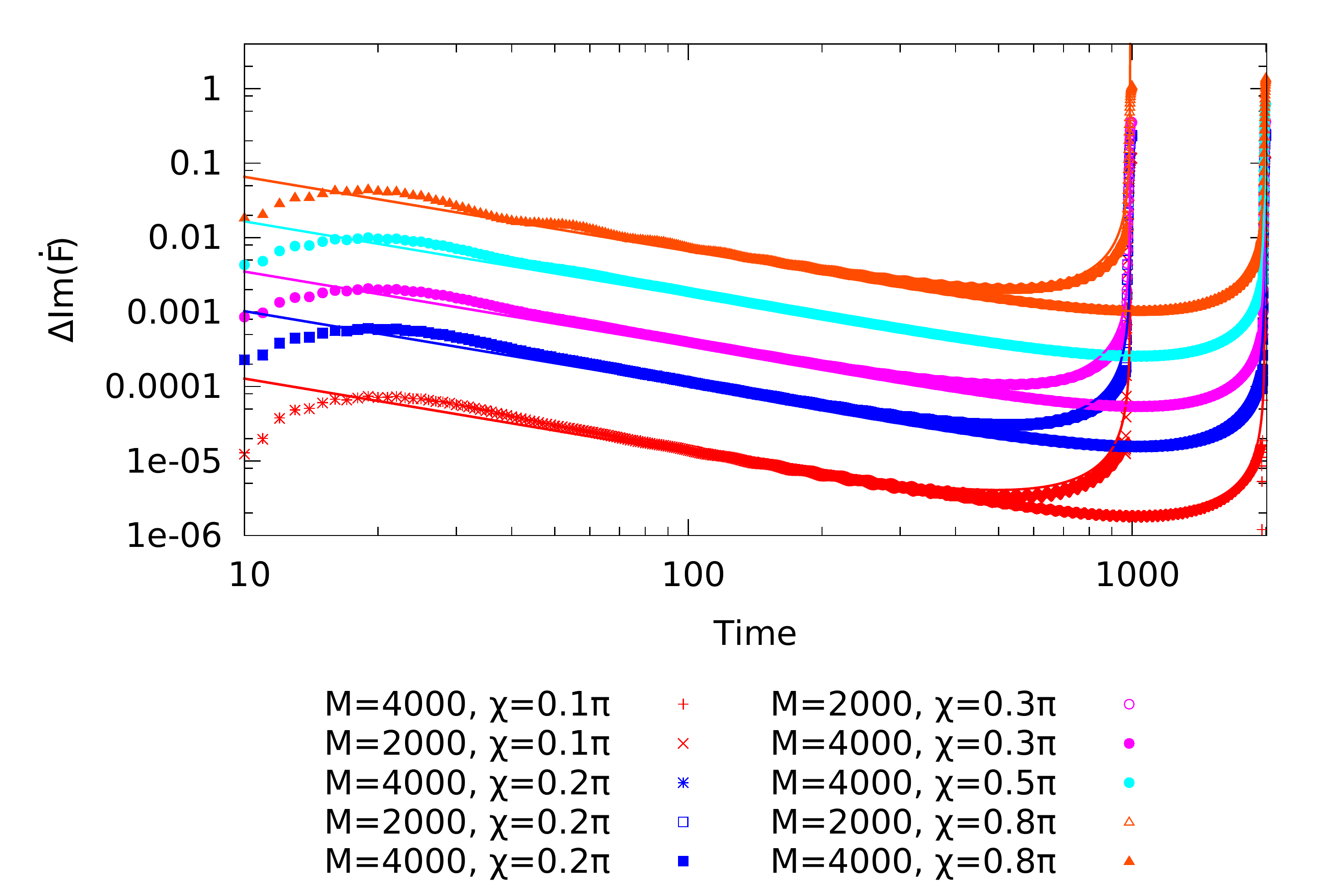}
\end{center}
\caption{Real part (left panel) and imaginary part (right panel) of the CGF for the RLM as a function of measuring time for various values of counting field $\chi$ and system size $M$,
         where the analytical infinite time result $\tilde{F}_{0}$ has been subtracted from the numerical result.
         The symbols are the results from real time numerical simulation, while the lines are from the analytic result \eqref{eq:series} with the conformal time substitution \eqref{eq:ConformalTime}, and $\tilde{F}_{0,1}$ given by Eqs.~\eqref{eq:F0-RLM} and \eqref{eq:F1-RLM}.  It is clearly seen that after an initial transient, the analytic formula fits the remainder of the data extremely well.}\label{fig:RLM-time-evolution}
\end{figure*}

The numerical data for the time evolution of the CGF is obtained using essentially the method of Ref.~\cite{CBS-2011}.
Specifically we use a symmetric version where one half of the counting field couples to the left contact and the other half to
the right contact. The overlap is calculated using the determinant formulas described in \cite{Schoenhammer-2007}.
Typical results as a function of measuring time, are shown in Fig.~\ref{fig:RLM-time-evolution} for various values of the counting field, as compared to analytic results from \eqref{eq:series} with the coefficients in Eqs.~\eqref{eq:F0-RLM}.
One can see that there are deviations from the $1/\tm$ behavior before the transit time is reached.  However the conformal time substitution \eqref{eq:ConformalTime}, which is used in these plots, captures this upturn perfectly.  In other words, after the transients, the first two terms of the series \eqref{eq:series} give a very good agreement over the whole of the steady-state region.

Fitting the real time numerical data to the ansatz  \eqref{eq:series} then gives numerically obtained  $\tilde{F}_0(\chi)$ and $\tilde{F}_1(\chi)$, which are plotted in Fig.~\ref{fig:dFxi}, along with the analytic results.
  As can be seen, the agreement is excellent, particularly in $\tilde{F}_1$ where the agreement is over six orders of magnitude.  We note that to get this level of agreement, it was important to use the analytic results above for the cosine band, and not to simply take the wide-band limit.  This may be an issue in the case of the IRLM, where only the wide-band (field-theoretic) result is known.
  
We also point out that the RLM is special when it comes to finite time effects -- the resonance condition means that in particular the noise (and higher cumulants) are very small at small bias voltage, while the finite time corrections may be much bigger.  This makes the RLM (and its interacting counterpart) an ideal model to test the theory of these finite time effects.

\begin{figure*}
\begin{center}
\includegraphics[width=0.49\columnwidth,clip=true]{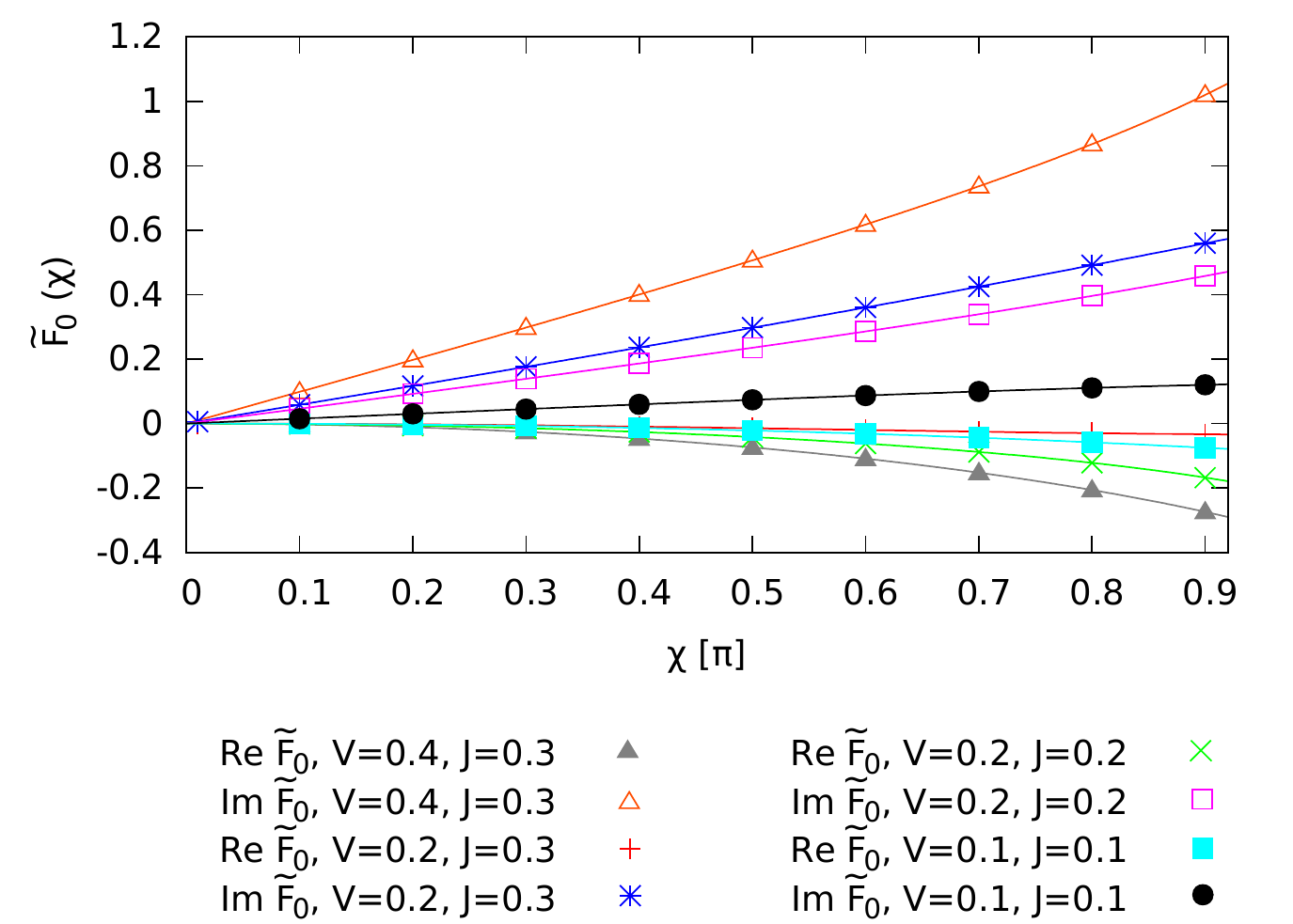}
\includegraphics[width=0.49\columnwidth,clip=true]{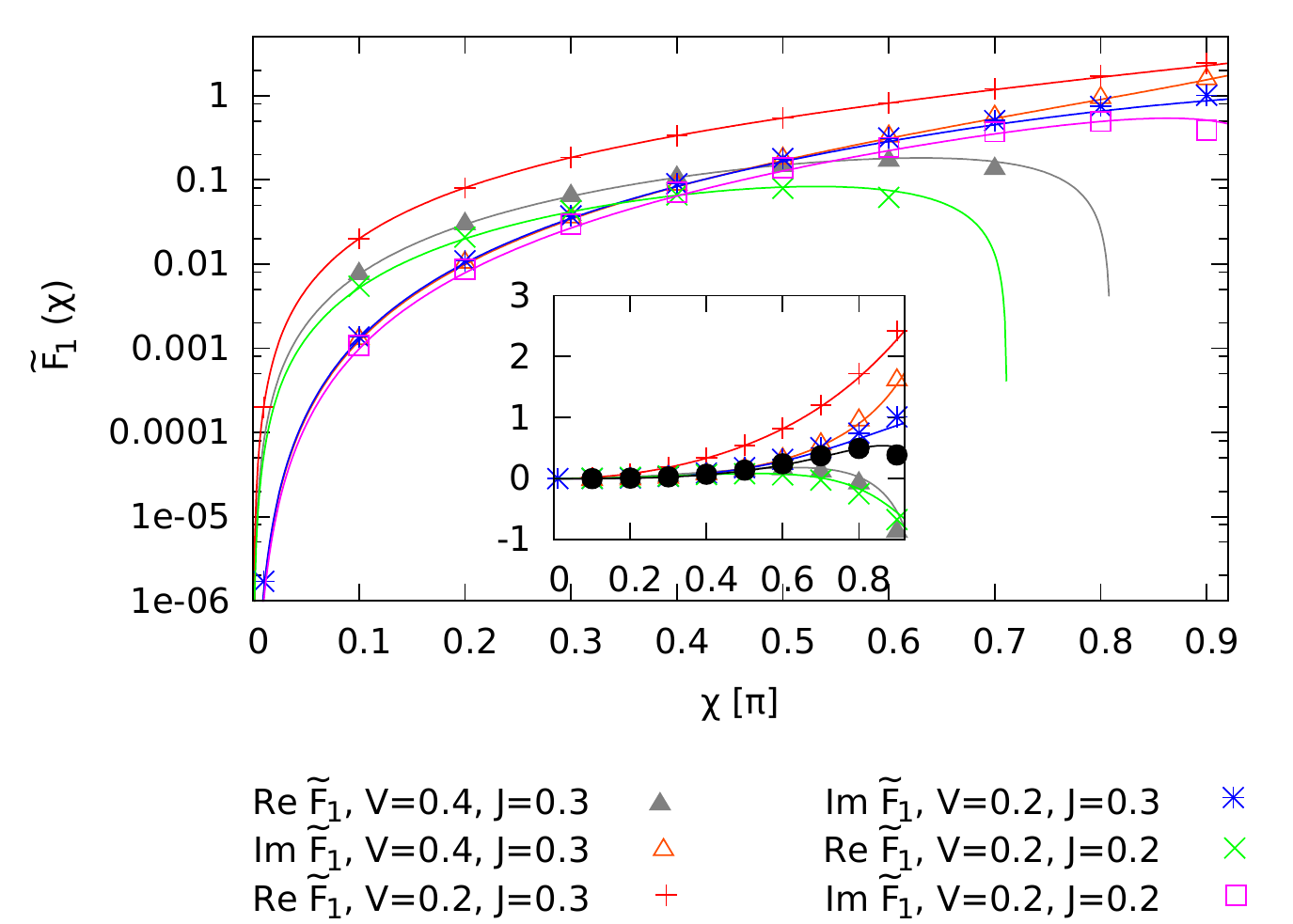}
\end{center}
\caption{Leading (left) and subleading (right) contributions to the full counting statistics of the non-interacting RLM (The inset gives the same result on a linear y axis.). Symbols correspond to the result obtained
   by fitting a $\tilde{F}_0 + \tilde{F}_1/d(\tm)$ behaviour to the time evolution in numerical simulation, as shown in Fig.~\ref{fig:RLM-time-evolution}. The lines give the analytical results given by Eqs.~\eqref{eq:F0-RLM} and \eqref{eq:F1-RLM}.}\label{fig:dFxi}
\end{figure*}

\subsection{$\tilde{F}_0$ and $\tilde{F}_1$ for the self-dual interacting resonant level model}

The analytic expression for the CGF of the self-dual interacting resonant level model, as derived from thermodynamic Bethe ansatz, is given by \cite{sm:CBS-2011}:
\be
\tilde{F}_0(\chi) = -\im V_{\mathrm{SD}} \chi - V_{\mathrm{SD}}\!\!\sum_{m > 0}  \frac{a_4(m)}{2m} \left( \frac{V_{\mathrm{SD}}}{T_B'} \right)^{6m} \left( e^{-2m\im\chi} -1 \right), \label{eq:F0-IRLM}
\ee
where
\be
a_K(m)=\frac{(-1)^{m+1} \sqrt{\pi} \, \Gamma \left(1+Km \right) }{2 m! \,\Gamma \left(\frac{3}{2} +(K-1)m\right) }.
\ee
and $T_B'\ = 2.7(J')^{4/3}$, the non-universal pre-factor $2.7$ relating the regularization of the field theory to the lattice model and is taken from previous work \cite{sm:Boulat-Saleur-Schmitteckert-2008}.  This means that there are no fitting parameters to compare analytic and numerical results, the plots simply show a direct comparison.

There are two points worth mentioning about the expression \eqref{eq:F0-IRLM}.  The first is that this series only converges for $V_{\mathrm{SD}}$ smaller than some critical value.  There is another series giving the expression for large $V_{\mathrm{SD}}$ (see \cite{sm:CBS-2011} for details), however for larger bias voltages it turns out that the finite time effects are much smaller, so we don't consider this limit here.  Secondly, the above expression is for the wide-band limit; while the numerics is done on a cosine (tight-binding lattice) band.  No exact results are known about the IRLM on a lattice.  However, it turns out that the discrepancy between these models is far less severe than for the non-interacting RLM.

Finally, for the sake of completeness, the expression for the sub-leading corrections which we compare to numerics is derived from Eq.~\ref{eq:F0-IRLM} via the conjecture \eqref{eq:F1}
\begin{equation}
\tilde{F}_1(\chi) = \frac{1}{\pi} \left[ \im \chi + \sum_{m > 0}  a_4(m)\frac{6m+1}{2m} \left( \frac{V_{\mathrm{SD}}}{T_B'} \right)^{6m} \left( e^{-2m\im\chi} -1 \right) \right]^2.  \label{eq:F1-IRLM}
\end{equation}
Comparing Eqs.~\ref{eq:F0-IRLM} and \ref{eq:F1-IRLM}, one sees that by far the most important finite time corrections appear in the real part of the CGF (see also the previous work \cite{sm:CBS-2011}).  This is why we focus only on the real part in the main text.

\end{widetext}

\end{document}